\documentclass[conference]{IEEEtran}
\IEEEoverridecommandlockouts
\usepackage{cite}
\usepackage{amsmath,amssymb,amsfonts}
\usepackage{algorithmic}
\usepackage[pdftex]{graphicx}
\usepackage{textcomp}
\usepackage{xcolor}
\usepackage[capitalise]{cleveref}
\usepackage{caption}
\usepackage{subcaption}
\usepackage[export]{adjustbox}
\usepackage{dblfloatfix}
\usepackage{multicol}
\usepackage{placeins}
\usepackage{balance}

\newcommand{\bseq}{\begin{subequations}}
	\newcommand{\eseq}{\end{subequations}}
\newcommand{\baln}{\begin{align}}
	\newcommand{\ealn}{\end{align}}
\newcommand{\balnd}{\begin{aligned}}
	\newcommand{\ealnd}{\end{aligned}}
\newcommand{\beq}{\begin{equation}}
	\newcommand{\eeq}{\end{equation}}
\newcommand{\beqn}{\begin{eqnarray}}
	\newcommand{\eeqn}{\end{eqnarray}}
\newcommand{\beqno}{\begin{eqnarray*}}
	\newcommand{\eeqno}{\end{eqnarray*}}
\newcommand{\bma}{\begin{displaymath}}
	\newcommand{\ema}{\end{displaymath}}
\newcommand{\bnu}{\begin{enumerate}}
	\newcommand{\enu}{\end{enumerate}}
\newcommand{\bce}{\begin{center}}
	\newcommand{\ece}{\end{center}}
\newcommand{\btb}{\begin{tabular}}
	\newcommand{\etb}{\end{tabular}}
\newcommand{\ba}{\begin{array}}
	\newcommand{\ea}{\end{array}}

\def\BibTeX{{\rm B\kern-.05em{\sc i\kern-.025em b}\kern-.08em
    T\kern-.1667em\lower.7ex\hbox{E}\kern-.125emX}}
\begin{document}

\title{Harnessing the Power of Swarm Satellite Networks with Wideband Distributed Beamforming \\
}

\author{\IEEEauthorblockN{Juan Carlos Merlano Duncan,~Vu Nguyen Ha,~Jevgenij Krivochiza, ~Rakesh Palisetty, ~Geoffrey Eappen, \\~Juan Andres Vasquez,~Wallace Alves Martins,~Symeon Chatzinotas, Björn Ottersten}
\IEEEauthorblockA{\textit{Interdisciplinary Centre for Security, Reliability and Trust, University of Luxembourg, Luxembourg}} 
}

\maketitle

\begin{abstract}

The space communications industry is challenged to develop a technology that can deliver broadband services to user terminals equipped with miniature antennas, such as handheld devices. One potential solution to establish links with ground users is the deployment of massive antennas in one single spacecraft. However, this is not cost-effective. 
Aligning with recent \emph{NewSpace} activities directed toward miniaturization, mass production, and a significant reduction in spacecraft launch costs, an alternative could be distributed beamforming from multiple satellites.
In this context, we propose a distributed beamforming  modeling technique for wideband signals.  We also consider the statistical behavior of the relative geometry of the swarm nodes. The paper assesses the proposed technique via computer simulations, providing interesting results on the beamforming gains in terms of power and the security of the communication against potential eavesdroppers at non-intended pointing angles. This approach paves the way for further exploration of wideband distributed beamforming from satellite swarms in several future communication applications.


\end{abstract}

\begin{IEEEkeywords}
SatCom, Swarm satellite, Wideband transmissions, Distributed beamforming.
\end{IEEEkeywords}

\section{Introduction}\label{sec:intro}
The rise in demand for broadband services offered through Satellite Communications (SatCom) has significantly increased the need for more efficient communication payloads. Furthermore, integrating satellite components into Terrestrial Networks (TNs) offers numerous opportunities for exploration. This integration can substantially augment the coverage of future cellular and Internet of Things (IoT) networks by incorporating a complementary non-TN. However, dealing with limited SatCom capabilities in smartphones and low-cost IoT terminals is challenging. These devices, equipped with small antennas, pose considerable constraints on link budgets \cite{Azari2022,VuHa_ICC23,Raka22}.

A viable strategy to navigate this challenge involves deploying spacecraft equipped with significantly larger antenna apertures \cite{VuHa_GCWks22,VuHaGC2022,VuHa_arxiv23, hayder_9852737}. Today's high-end spacecraft, typically weighing between 500 and 1500~kg, often incorporate monolithic antennas with apertures covering a few square meters~\cite{LAREQ}. However, the costs associated with designing, constructing, and launching such sizable satellites can be prohibitive. To satisfy the demanding requirements for broadband coverage in User Equipment (UE) with limited SatCom capabilities, the antenna apertures of current spacecraft need to be increased by at least tenfold, expanding to hundreds of square meters~\cite{LAREQ}. 

Nevertheless, augmenting the size of monolithic antenna structures requires careful consideration of microgravity forces in operation and substantial accelerations experienced during satellite launches~\cite{LAREQ}. In essence, deploying larger monolithic satellites results in a significant increase in weight and costs. Thus, the deployment of giant payload antennas is a high-risk undertaking, as demonstrated by the well-known failure of the Solaris Mobile project. However, regarding beamforming capabilities, a constellation of small satellites could replace an advanced Low Earth Orbit (LEO) satellite weighing around 1000~kg, each weighing 100~kg or less~\cite{LAREQ}.

\begin{figure}[t!]
    \centering
    \includegraphics[width=\linewidth]{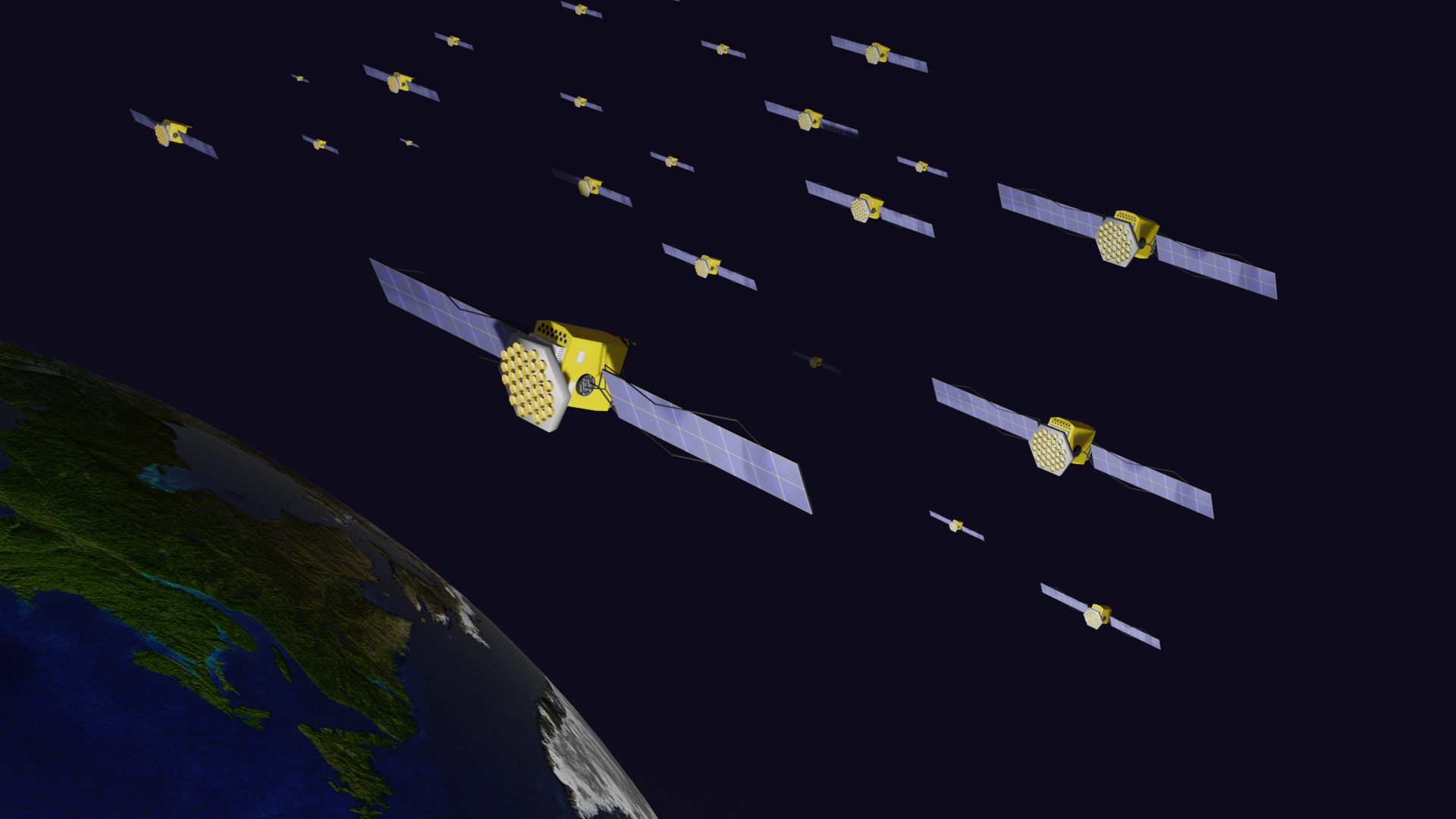}
    \captionsetup{font=footnotesize}
    \caption{Small communication satellites flying in close formation. Each satellite is equipped with a small planar antenna array}
    \label{fig:Art1}
\end{figure}


Satellite swarm networks are poised to play a pivotal role in future communication systems, particularly in providing ubiquitous connectivity and facilitating new applications. Nonetheless, such design and operation present many challenges and necessitate innovative solutions. In \cite{9939157}, the authors introduce a novel LEO Satellite Network (SatNet) architecture, leveraging Distributed Massive Multiple-Input Multiple-Output (DM-MIMO) technology that enables direct communication between satellites and mobile devices. In \cite{Knopp2023}, a new LEO SatNet architecture is proposed, utilizing swarm-based beamforming/signal-combining to achieve superior performance and cost efficiency compared to conventional single satellite systems. The work in~\cite{8293791} reviews the recent advancements in formation control of small satellites—groups of satellites that fly in coordinated patterns to serve various applications. The work in~\cite{BACCI_icssc22} examines the use of several satellites with regular planar arrays arranged in a square formation. These works collectively provide a comprehensive overview of the current state-of-the-art and prospects in SatCom swarm research. 

The current trend of the space economy, commonly referred to as \emph{NewSpace}, is centered around the miniaturization, mass production, and enhanced capabilities of spacecraft~\cite{Kodheli2021}. Rather than launching a single large monolithic spacecraft, this approach involves deploying numerous small spacecraft, some weighing only kilograms. These smaller spacecraft leverage advancements in software capabilities and electronics miniaturization. This shift in space system design introduces a new paradigm by incorporating redundancy through multiple spacecraft and reducing capital expenditures through cost spread over time~\cite{culbert2022}. However, it also presents unique challenges. The physical limitations imposed by their small size, such as antenna aperture size, constrain their communication capabilities, power generation, orbit, and attitude control efficiency~\cite{Kodheli2021, Liz9761868}. One possible solution to overcome SatCom's limitations is to utilize a swarm of small spacecraft that collectively function as a non-monolithic satellite, employing an immense Virtual Antenna Array (VAA)~\cite{Dohler2004}. However, the design scenario differs from conventional antenna arrays, and the commonly known concepts from phase arrays do not hold. 
 
In this paper, we hypothesize that the only way to analyze a swarm array with such a large baseline compared to the central carrier frequency wavelength is to eliminate the narrowband assumption and completely adopt a wideband model. The "baseline" refers to the maximum separation between the nodes, borrowing the definition from interferometry and radio astronomy \cite{baseline_def}.  For instance, the narrowband assumption of the phased array presented in classical array theory is inadequate to model the swarm accurately. Similarly, evaluating grating lobes at a particular exact frequency does not provide conclusive information to evaluate the array performance. Works in~\cite{Aksoy2014, Afacan2012} corroborate this claim and analyze the effect of shifting pulses of a Time-Modulated Array (TMA) for asymmetric volumetric arrays.

For the wideband modeling, a handful---e.g., tens or hundreds---of small LEO communication satellites would fly in relatively close formation, or `swarm,' as depicted in Fig.~\ref{fig:Art1}. These satellites would cooperate—exchanging data and control signaling—to communicate with a target through coherent processing across the swarm. In this context, we posit that `Distributed Beamforming' (DBF) is the crucial technological empowerer to meet these stringent power-budget requirements.  We propose to use the randomness of the spacecraft position to the designer's advantage and to relax the requirements related to the spacecraft’s physical limitations and flight dynamics affecting the swarm. These behaviors depend on the accuracy and miniaturization of position-keeping systems, which cannot maintain a fixed relative geometry of the swarm under realistic conditions.  Thus, we model the wideband beamforming using baseband Nyquist pulses in addition to the propagation phases. In the model, a time-domain pulse, for example, a sinc or square-root-raised-cosine filter, modulates the main carrier. 
Finally, we evaluate this model numerically using high-level computer simulations. 

The rest of this paper is organized as follows: Section~\ref{sec:model} describes the wideband beamforming model, including the models for the satellite geometrical distribution.  Section~\ref{sec:discrete} presents an equivalent discrete-time model for the beamforming evaluation. Section~\ref{sec:evaluation} presents the numerical results of the wideband beamforming operation, and Section~\ref{sec:conclusions} concludes the paper.

\section{SatCom Swarm System Model}
\label{sec:model}
\begin{figure}[t!]
    \centering
    \includegraphics[width=\linewidth]{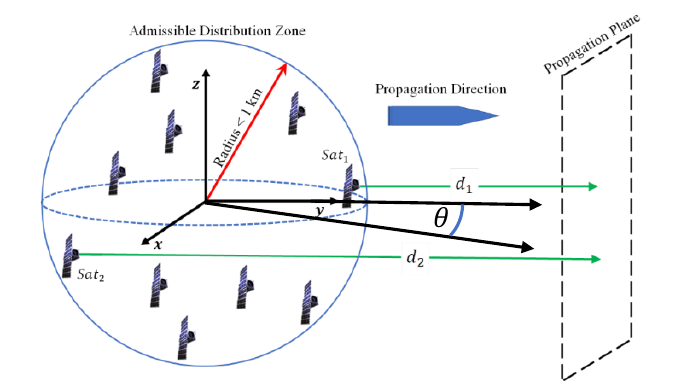}
    \captionsetup{font=footnotesize}
    \caption{Simplified geometry for the satellite swarm. }
    \label{fig:geometry}
\end{figure}

This section aims to present a model of the SatCom swarm to evaluate the system's performance. The geometry of the SatCom swarm is depicted in Fig.~\ref{fig:geometry}. In this model, $N$ satellites, also referred to as nodes, fly in tight formation in either an LEO or Medium Earth Orbit (MEO). The swarm's primary objective is to transmit a signal to a single user located on the ground, although this model can be extended to accommodate multiple users.

The model assumes that each node position is entirely random, and the nodes are kept in a virtual spherical bound. This limit is set to a 1 km radius, just as illustrated in the figure. We model the relative position of a node $n \in \{1,\dots,N\}$ as a random vector $ \boldsymbol{p}_n$ in the 3D Euclidean space.  The components of  the random vector $\boldsymbol{p}_n$ are independent and identically distributed (iid), with zero-mean (i.e., the mean of $\boldsymbol{p}_n$ is the center of the swarm bounding sphere) and variance  $\sigma_{\boldsymbol{p}}^2$.  We assume that $\boldsymbol{p}_n$ maintains constant during a coherence time, which we use for further calculations.

Let us start by establishing a reference pointing angle. This angle is critical as it represents the core direction toward which the swarm is programmed to direct its transmissions. We have set this reference angle to align with the $y$-axis, an intentional design decision owing to the $y$-axis's orientation toward the Earth's center. Therefore, the unit vector $\boldsymbol{u}_y = \begin{bmatrix}0& 1& 0\end{bmatrix}^{\rm T}$ is critical in defining a reference propagation plane, employing a reference distance denoted as $d$. This distance is an essential parameter for our calculations. However, for simplicity in our computations and reasoning, we can assume that the distance $d$ corresponds to the shortest possible distance to the nearest point on Earth.
However, the swarm is not rigidly fixed to only one pointing direction. It has been designed to direct its signals toward different angles, denoted explicitly as $\theta$ and $\phi$ in our model. For the sake of clarity, Fig.~\ref{fig:geometry} only displays $\theta$ and purposely omits $\phi$.

To represent the new tilted angle to which the swarm may direct its transmissions, we have to express this as a vector. This new vector is essentially a rotated version of our reference pointing vector. The applied rotation is determined by the rotation angles $\theta$ and $\phi$, as follows:
\begin{align}\label{eq:RotationMatrix}
\boldsymbol{u}_{\theta,\phi} 
&=\underbrace{\begin{bmatrix}
\cos \left(\theta \right) & -\cos \left(\phi \right)\sin \left(\theta \right) & \sin \left(\phi \right)\sin \left(\theta \right)\\
\sin \left(\theta \right) & \cos \left(\phi \right)\cos \left(\theta \right) & -\cos \left(\theta \right)\sin \left(\phi \right)\\
0 & \sin \left(\phi \right) & \cos \left(\phi \right)
\end{bmatrix}}_{= \boldsymbol{\Upsilon}}\boldsymbol{u}_y \nonumber\\
&= \boldsymbol{\Upsilon} \boldsymbol{u}_y \,.
\end{align}

This section aims to calculate the distance from any individual satellite located at $\boldsymbol{p}_n$ to a given tilted plane whose distance to the origin is again $d$. This is a crucial calculation for understanding the satellite's relative positioning in space. This distance, denoted as $d_n$, is
\beq d_n = d - \boldsymbol{p}_n^{\rm T}\,\boldsymbol{u}_{\theta,\phi}\,. \eeq
The inner product $\boldsymbol{p}_n^{\rm T}\,\boldsymbol{u}_{\theta,\phi}$ calculates the projection of the satellite position vector onto the direction of the normal plane. Subtracting this projection from the distance $d$ yields the desired distance from the satellite to the plane, $d_n$.

Next, we want to determine the difference in distances from the satellite to two distinct planes. This calculation provides important information about the satellite's relative position with respect to these planes, which is useful for trajectory analysis and navigation. The range difference is given by
\begin{align}
\Delta_{n} & = ( d - \boldsymbol{p}_n^{\rm T}\,\boldsymbol{u}_{\theta,\phi}) - (d - \boldsymbol{p}_n^{\rm T}\,\boldsymbol{u}_{y} ) \nonumber \\
& = \boldsymbol{p}_n^{\rm T}(\boldsymbol{I} -  \boldsymbol{\Upsilon}  ) \boldsymbol{u}_{y}\,,
\end{align}
where $\boldsymbol{I}$ is the $3\times 3$ identity matrix.

The objective is to combine the signals of $N$ satellites in an intended pointing direction.  We can calculate this combination on both the reference and tilted planes, with the key assumption that the inter-satellite distance is significantly smaller than the propagation distance $d$.
Under this assumption, and considering that each satellite transmits the signal $s_n(t)$ using an omnidirectional\footnote{Herein, we made this omnidirectionality assumption for the sake of clarity, without loss of generality.} radiator, then the signal impinging upon the tilted propagation plane can be written as
\begin{align}
r(t) =  \sum_{n=1}^N{ s_n\left( t - \tau + \tilde{\tau}_n   \right) } {\rm e}^{-{\rm j} \Theta_n },
\end{align}
in which $\tau = d/c$ and $\tilde{\tau}_n = (\boldsymbol{p}_n^{\rm T}\,\boldsymbol{u}_{\theta,\phi})/c$, with $c$ denoting the speed of light,  and 
\begin{align}
    \Theta_n= 2\pi\left( - \tau + \tilde{\tau}_n  \right)  f
\end{align}
where $f$ is the central carrier frequency. We can verify  by simple inspection that the combination might be that the expected maximum bandwidth for narrowband signals becomes in the orders of tens of kilohertz when the variability in  $\sigma_{\boldsymbol{p}}$ is in the order of hundreds of meters.

Let us consider a scenario where the swarm is tasked with transmitting the signal $v(t)$ to a specific user on the ground. The method used to achieve this is through precompensated signals, expressed as
\begin{equation}
s_n(t) = v(t - \tilde{\tau}_n) {\rm e}^{{\rm j} \Theta_n }\,.
\label{eq:complexpulse}
\end{equation}
The precompensation time values $\tilde{\tau}_n$ are of paramount importance for ensuring that the transmission from each satellite in the swarm arrives at the ground user at the correct time to achieve an effective signal combination.
However, it is essential to note that signal transmission is not perfectly directional. Thus, signals are also perceived in non-intended directions. These unintended signals can be written as 
\begin{equation}
 r(t) =   \sum_{n=1}^N{ s_n \left( t - \tau + \frac{ \Delta_{n} }{c}   \right) } {\rm e}^{ - {\rm j} 2 \pi \frac{\Delta_{n}}{c}f}\,.  
\end{equation}

In the specific instance of the reference propagation direction denoted by $\boldsymbol{u}_y$, the random variable $\Delta_{n}$ depends on the second column of the rotation matrix, as highlighted in~\eqref{eq:RotationMatrix}. By evaluating this, we derive the variance of this propagation difference, represented by
\begin{align}
  \sigma^2_{\Delta_n} &=   \left((\cos \phi \sin \theta )^2 + (1 -  \cos \phi \cos \theta )^2 + ( \sin \phi )^2
\right) \sigma_{\boldsymbol{p}}^2 \nonumber\\
&= 2(1-\cos \phi \cos \theta)\sigma_{\boldsymbol{p}}^2\,.
\end{align}
This equation encapsulates the variability of the propagation difference $\Delta_{n}$ due to the spatial orientation of each satellite in the swarm. 


\section{Discrete-Time model}
\label{sec:discrete}
 We assume that the entries of the $L\times 1$ vector $\boldsymbol{v}$ are the discrete-time series of transmitted complex-valued symbols, which are iid, with zero mean and unitary variance. Further, we assume that the block length $L$ is sufficiently long, so that it is much larger than a pulse duration. 
 With these assumptions, we can model the transfer response as a multiplication of a matrix by a vector, where the matrix  $\boldsymbol{Q}$, is a circulant matrix. The first row of this matrix consists of the discretized samples of the complex-valued Nyquist pulse resulting from the combination process. 
Under these conditions, the signal received by a target receiver, assuming free-space losses have been normalized, can be represented as follows:
 \begin{equation}
     \boldsymbol{r} = \boldsymbol{Q}\cdot\boldsymbol{v}\,.
 \end{equation}
 
Our discrete-time analysis incorporates the translation of the continuous delays by deploying equivalent linear-phase delay filters, as outlined in \cite{231921}. This approach takes into account the summation of all complex-valued pulses, each with their corresponding phase offset, as expressed in~\eqref{eq:complexpulse}.

In this context, the complex-valued pulses represent the effect of individual satellite transmissions. These are not isolated events; instead, they exist in a continuous stream of data where each transmission affects and is affected by those around it. The inclusion of phase offset in the summation allows for the consideration of how each pulse is adjusted in time, effectively capturing the varying delay experienced by signals depending on their path to the receiver.
By leveraging this analytic method, we can construct a comprehensive model of the total received signal power. This model is not only descriptive, revealing the interplay of variables and their collective influence on the received signal, but it is also predictive, enabling expectations about system performance under various conditions. The use of linear-phase delay filters and complex pulse summation provides a robust, nuanced approach to understanding and predicting the behavior of the system.
Consequently, we can express the expected raw power of $\boldsymbol{r}$ as
\begin{align}
    P_{\boldsymbol{rr}} &= \frac{1}{L}\cdot\mathrm{E}\left\{
  \mathrm{trace}(\boldsymbol{r}\cdot\boldsymbol{r}^{\rm H})
\right\}\nonumber\\ 
&= 
\frac{1}{L}\cdot\mathrm{trace}(\boldsymbol{Q}\boldsymbol{Q}^{\rm H})\,.
\label{eq:prr}
\end{align}

Following the computation of the total raw power, we introduce an estimator to assess the similarity of the recovered signal, $\tilde{\boldsymbol{r}}$, to the original transmitted symbol, $\boldsymbol{v}$. The recovered signal is given by $\tilde{\boldsymbol{r}} = \boldsymbol{Q}_{\rm i}\boldsymbol{Q}\boldsymbol{v}$, where $\boldsymbol{Q}_{\rm i}$ represents the circulant matrix of the matched Nyquist pulse under ideal conditions. The estimator essentially gauges the quality of the recovered symbol. 
It does so by correlating the recovered signal with the ideal matched pulse as follows: 
\begin{align}
P_{\boldsymbol{v}\tilde{\boldsymbol{r}}}  &=  \frac{1}{L}\cdot\mathrm{E}\left\{
  \mathrm{trace}(\boldsymbol{v}\cdot\tilde{\boldsymbol{r}}^{\rm H})
\right\} \nonumber\\
&= \frac{1}{L}\cdot\mathrm{E}\left\{
  \mathrm{trace}\left(\boldsymbol{v}\cdot(\boldsymbol{Q}_{\rm i}\boldsymbol{Q}\boldsymbol{v})^{\rm H}\right)
\right\}\nonumber\\
&= \frac{1}{L}\cdot\mathrm{trace}\left((\boldsymbol{Q}_{\rm i}\boldsymbol{Q})^{\rm H}\right),
\label{eq:prv}
\end{align}
Given that the matrix $\boldsymbol{Q}$ remains constant for a single channel realization, the value of $P_{\boldsymbol{v}\tilde{\boldsymbol{r}}}$ effectively becomes the dot product of the combined pulse and the ideal matched filter. This results in the maximum value at the intended symbol time instant. Importantly, this estimator offers more than just a measure of signal recovery quality. It provides a means of quantifying the security of the communication against potential eavesdroppers at non-intended pointing angles. In other words, it helps evaluate the robustness of the system in maintaining the confidentiality of the transmitted information. By providing a peak at the intended symbol time and maintaining low values elsewhere, the system inherently offers a degree of resistance against unauthorized interception of the signal.

Building on this mathematical framework, we then delve into the subsequent section, which presents a comprehensive numerical evaluation of the statistical parameters. This numerical assessment focuses on the gains accrued from the signal combination process (beamforming) at varying pointing angles. These evaluations will provide essential insights into how different pointing strategies can impact the overall system performance in terms of signal strength and quality.

\section{Numerical Assessment of Beamforming Gains}
\label{sec:evaluation}
This section evaluates the signal combination performance for swarms of different numbers of nodes and multiple position variances.  The assessment is conducted in the discrete-time domain with specific carrier frequency and baud rate values, as shown in Table~\ref{setting_paras}, which summarizes all the simulation parameters. The array gain is averaged with multiple distance realizations of the swarm distributions. We assume omnidirectional antennas and disregard the free-space losses for simplicity. 

\begin{table}[!t]
	\centering
	\caption{\textsc{Simulation parameters}}
 \vspace{-2mm}
	\begin{tabular}{l| r}
		\hline
		Parameter & Value \\
		\hline\hline
		Pulse Bandwidth										& $50,500$~MHz \\
		Pulse Shape 	& SRRC\\
		Roll-off Factor & $0.2$ \\
		Oversampling Factor & $4$ \\
		Carrier Freq. & $20$~GHz \\
        Position distribution &  Gaussian \\
		Position Std-Dev $\sigma_{\boldsymbol{p}}$ & $500$, $1000$~m \\
  		Number of Nodes 	& $8,16,256$ \\ 
		Transmit Antenna					& Omnidirectional\\	
		Number of repetitions for averaging					& $200$ \\
		Number of SRRC samples												& $41$\\
		Number of re-sampling samples	(Oversampling)					& $4$\\
		Re-sampling order							&  Cubic   \\
	\hline
	\end{tabular} \label{setting_paras} 
	\end{table}
 
\begin{figure}[t!]
    \centering
    \includegraphics[width=\linewidth]{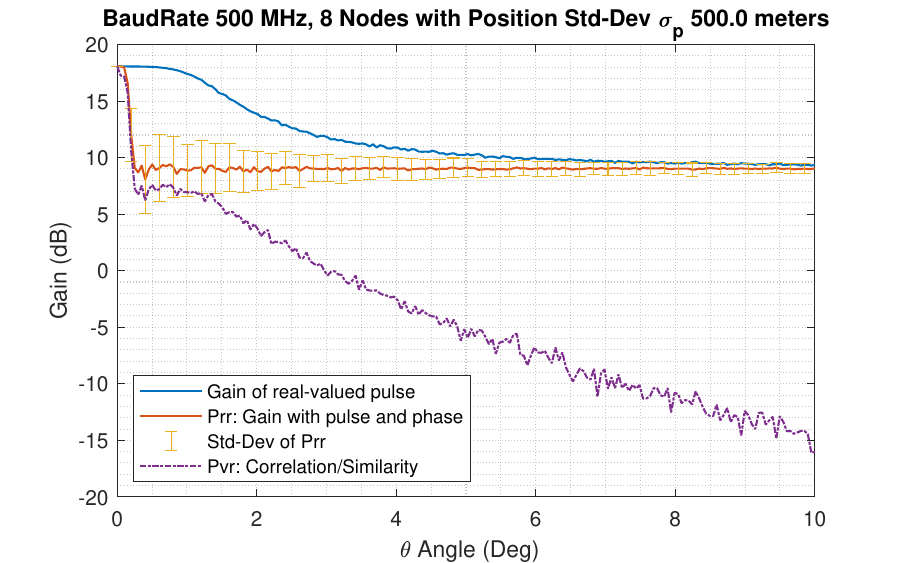}
    \captionsetup{font=footnotesize}
    \caption{Average expected Gains for multiple realizations of the array distribution. Baud Rate 500~MHz, 8 Nodes, $\sigma_p = 500$ m. (200 realizations) }
    \label{fig:Gain1}
\end{figure}

\begin{figure}[t!]
    \centering
    \includegraphics[width=\linewidth]{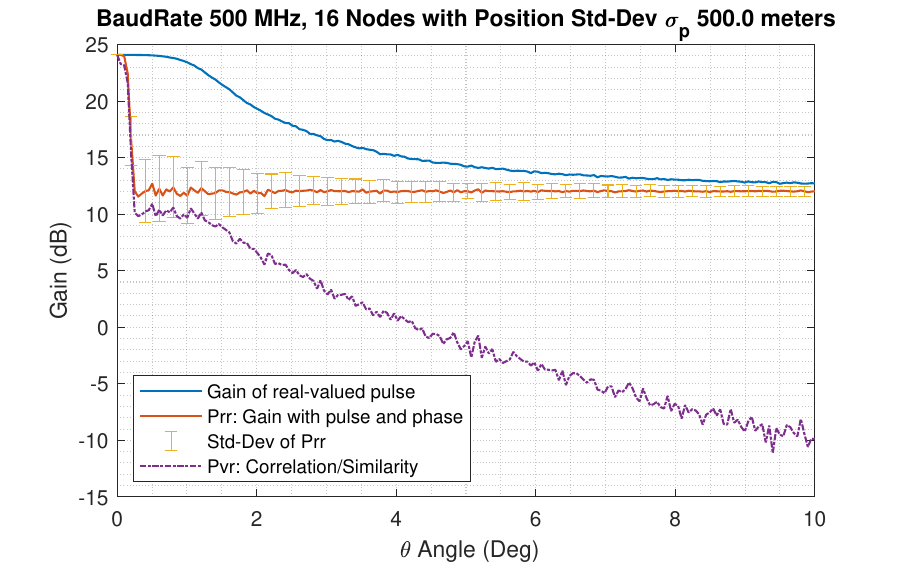}
    \captionsetup{font=footnotesize}
    \caption{Average expected Gains for multiple realizations of the array distribution. Baud Rate 500~MHz, 16 Nodes, $\sigma_{\boldsymbol{p}}$ = 500 m. (200 realizations) }
    \label{fig:Gain2}
\end{figure}

\begin{figure}[t!]
    \centering
    \includegraphics[width=\linewidth]{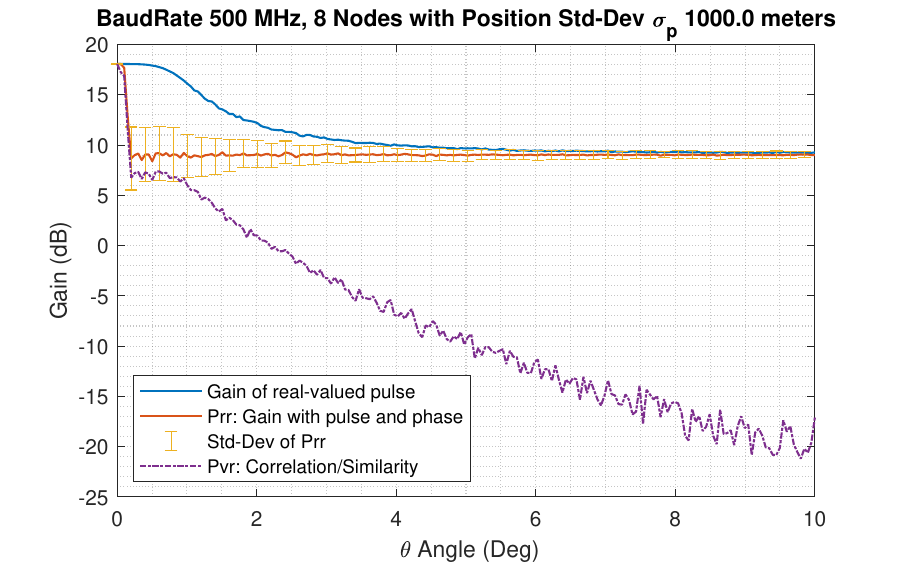}
    \captionsetup{font=footnotesize}
    \caption{Average expected Gains for multiple realizations of the array distribution. Baud Rate 500~MHz, 8 Nodes, $\sigma_{\boldsymbol{p}}$ = 1000 m. (200 realizations) }
    \label{fig:Gain3}
\end{figure}

\begin{figure}[t!]
    \centering
    \includegraphics[width=\linewidth]{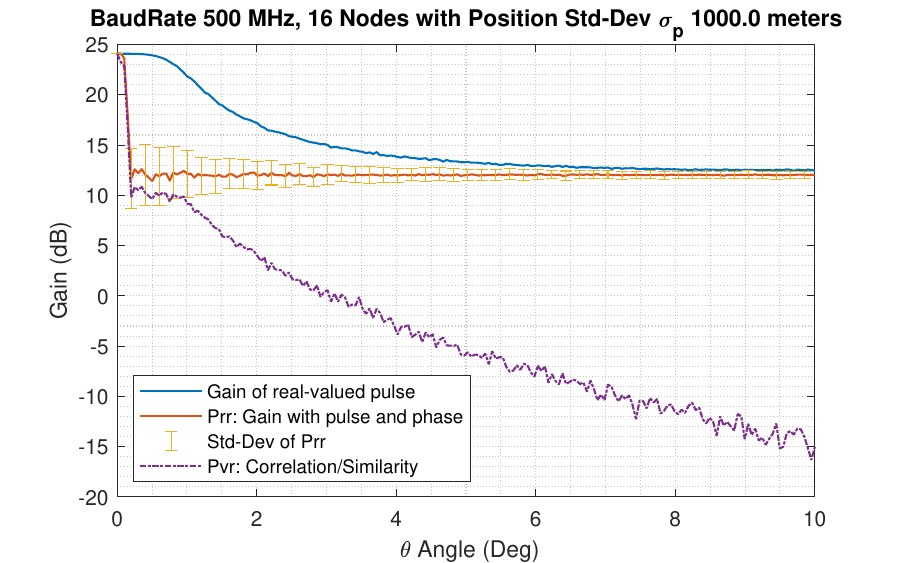}
    \captionsetup{font=footnotesize}
    \caption{Average expected Gains for multiple realizations of the array distribution. Baud Rate 500~MHz, 16 Nodes, $\sigma_p $= 1000 m. (200 realizations) }
    \label{fig:Gain4}
\end{figure}

In Figs.~\ref{fig:Gain1}-\ref{fig:Gain4}, we present a detailed analysis of the total received power over various values of the pointing angle, $\theta$. 
These simulation results demonstrate the relationship between angular gains and the pulse-shaping response.  Specifically, these beamforming gains discard the path phase $\Theta_n$ for varying values of $\theta$. These gains, depicted as blue curves and labeled as ``Gain of real-valued pulse", exhibit a decreasing trend as the pointing angle $\theta$ enlarges.
Furthermore, we have represented the averaged pulse energy according to $P_{ \boldsymbol{rr}}$ given in~\eqref{eq:prr}. These values averaged over $200$ realizations, are plotted as orange curves, providing a statistical perspective on the overall power performance.
To offer a more comprehensive view of power gains, we have also incorporated an estimation of the standard deviation of the $P_{ \boldsymbol{rr}}$ gains in these figures, which are illustrated in yellow.
Interestingly, $P_{ \boldsymbol{rr}}$ drops quickly as  $\theta$ takes a positive value, then saturates when $\theta \geq 1$~degree.
Additionally, we introduce the squared cross-correlation/similarity measurement from \eqref{eq:prv}, represented by the dashed purple curves. 
This inclusion further emphasizes the interactions between these variables and their collective impact on the total received power.
Like the beamforming gain, the squared cross-correlation and similarity measurement display a decreasing trend as the value of $\theta$ increases.


\begin{figure}[t!]
    \centering
    \includegraphics[width=\linewidth]{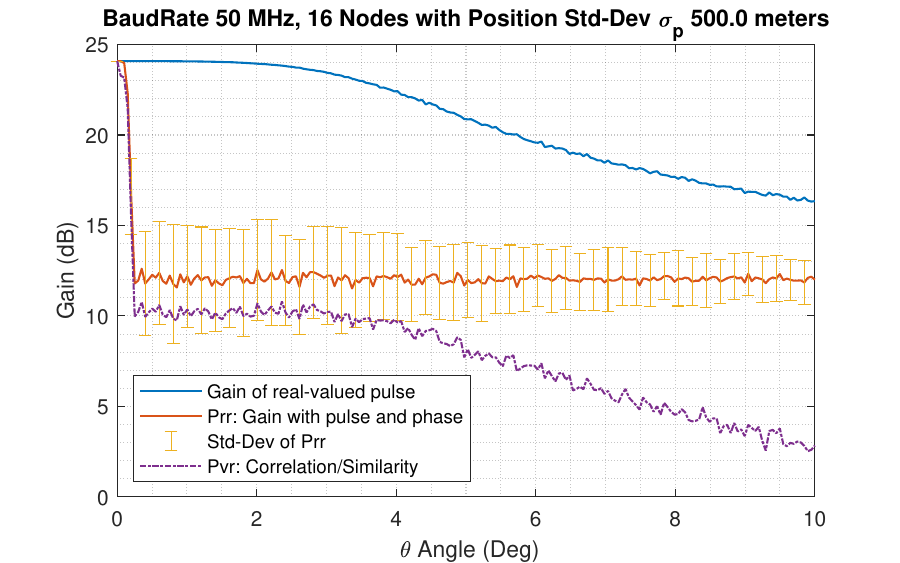}
    \captionsetup{font=footnotesize}
    \caption{Average expected Gains for multiple realizations of the array distribution. Baud Rate \textbf{50~MHz}, 16 Nodes, $\sigma_{\boldsymbol{p}}$ = 500 m.  (200 realizations) }
    \label{fig:Gain50M}
\end{figure}

\begin{figure}[t!]
    \centering
    \includegraphics[width=\linewidth]{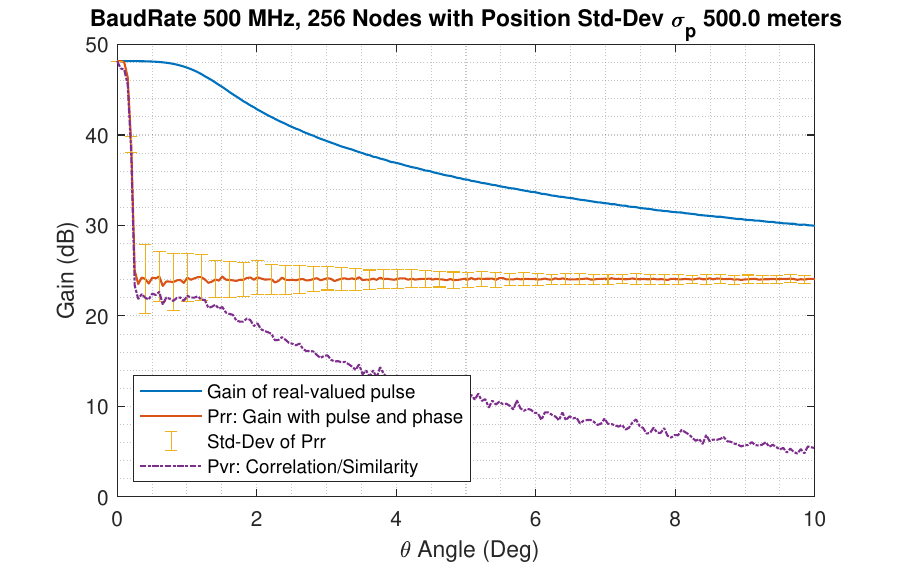}
    \captionsetup{font=footnotesize}
    \caption{Average expected Gains for multiple realizations of the array distribution. Baud Rate 500~MHz, \textbf{ 256 Nodes}, $\sigma_{\boldsymbol{p}}$ = 500 m. (200 realizations) }
    \label{fig:Gain6}
\end{figure}
Fig.~\ref{fig:Gain50M} shows the particular case when the signal baud rate is reduced to 50~MHz instead of 500~MHz, whereas Fig.~\ref{fig:Gain6} shows the expected gains of the scheme with 256 satellites (Nodes).
These figures demonstrate how the main beamwidth and gains behave as a function of various system parameters. The beamforming gain reaches $N^2$ at the beam center, an effect attributed to each new node adding power to the antenna gain during the beamforming process, as explained in \cite{Mudumbai_dbf}.
The expected beamwidth, in terms of $P_{ \boldsymbol{rr}}$ and $P_{ \boldsymbol{v \tilde{r} }}$, appears to be influenced primarily by the variance of the satellite position or, equivalently, the parameter $\sigma_{\boldsymbol{p}}$. The larger the $\sigma_{\boldsymbol{p}}$ value, the narrower the beamwidth becomes.
The out-of-beam gain of the $P_{ \boldsymbol{rr}}$ expected value reduces to $N$, which is $N$ times smaller than the gain in the main direction. The received signal correlation governs the angular signal selectivity, which narrows as the bandwidth increases.

\section{Conclusion and future challenges}
\label{sec:conclusions}


In conclusion, this work presented the vision of distributed beamforming from satellite swarms for enabling high data rate links with SatCom-limited UEs. We proposed a wideband model for distributed beamforming, taking into account the random distribution of satellite nodes.  The paper also elucidated the behavior of the main beamwidth and its gain as function of system parameters. In this sense, we challenged the suitability of the widespread belief, in the field of antennas, that there is a one-to-one relationship between gain and directivity.  Notwithstanding, several publications focusing on sparse and long-baseline arrays have noticed this effect.

We elaborated the system analysis by including a discrete-time equivalent, and finally, we validated it using computer-based numerical simulations for a carrier in the Ka-band. Our simulation results showed that, with each new node added to the array distribution, the signal power cumulatively enhances the gain of the onboard phase array. Therefore, combining signal power from multiple satellites can improve the gain at the UE. This process is coupled with efficient power generation, as onboard power generation proves to be more efficient when more satellites are employed. 
On the other hand, the statistics of the maximum array separation determine the angular directivity. As the maximum separation between any two nodes in the swarm increases, the beam becomes narrower.  As a salient consequence, the carrier phase synchronization requirement between the nodes becomes more strict and, in some cases, prohibitive. One can guess the solution would be restricting the satellites to a smaller volume. However, maintaining a tightly-closed formation of satellite swarms, with a standard deviation of $500$ meters, as seen in the simulation examples, remains a significant challenge. 
The simulation results do not include the effects of the antenna gains at each of the satellites. In practical scenarios, the swarm gains will add up on top of the satellite antenna gains, reducing the power radiation in unintended directions.
Finally, our approach is particularly applicable when multiple concurrent beams are transmitted, while a  research question remains regarding the computational complexity of the multi-beam beamforming operation. This process includes the data exchange between nodes and the coefficient-multiplication operation for all the transmitted signals. One possible alternative is the butterfly signal routing among the nodes, which is the building block of Fast Fourier Transform (FFT) algorithms. This topology might reduce the  number of interconnections among the nodes and the  number of multiplication operations~\cite{Raka22}.







\section*{Acknowledgement}
This work was supported by the Luxembourg National Research Fund (FNR), through the CORE Project (ARMMONY): ``\textit{Ground-based distributed beamforming harmonization for the integration of satellite and Terrestrial networks}'', under Grant FNR16352790, by the European Space Agency (ESA) under the project number 4000134678/21/UK/AL "\textit{User Terminal With Path Diversity For Constellations} (DIVERSITY)," and SES S.A. (Opinions, interpretations, recommendations and conclusions presented in this paper are those of the authors and are not necessarily endorsed by FNR, ESA or SES).

\bibliographystyle{IEEEtran}




\end{document}